\documentclass[prc,twocolumn,floatfix,groupedaddress,nofootinbib,preprintnumbers,
amsmath,amssymb,amsfonts,superscriptaddress,widetable] {revtex4}
\usepackage{bm}
\usepackage{mathrsfs}
\usepackage{amssymb}
\usepackage{amsmath}
\usepackage{graphicx}
\usepackage{array}
\usepackage{color}
\usepackage{tabularx}
\newcommand{\Msun}{\,M_{\odot}}

\def\apjl{Astrophys. Journal Lett.}
\def\physrep{Phys. Reports}
\def\aap{Astronomy \& Astrophysics}
\def\mnras{Monthly Notices of the Royal Astronomical Society}
\begin{document}
\title{Total Energy in Supernova Neutrinos and the Tidal Deformability and Binding Energy of Neutron Stars}


\author{Brendan Reed}\email{reedbr@iu.edu}
\affiliation{Department of Astronomy, Indiana University,
                  Bloomington, IN 47405, USA}
\affiliation{Center for Exploration of Energy and Matter and
                  Department of Physics, Indiana University,
                  Bloomington, IN 47405, USA}
\author{C. J. Horowitz}\email{horowit@indiana.edu}
\affiliation{Center for Exploration of Energy and Matter and
                  Department of Physics, Indiana University,
                  Bloomington, IN 47405, USA}
\date{\today}
\begin{abstract}
The energy radiated in supernova neutrinos is a fundamental quantity that is closely related to the gravitational binding energy of a neutron star.  Recently the tidal deformability of neutron stars was constrained by gravitational wave observations.  By considering several equations of state, we find a strong correlation between the tidal deformability and neutron star binding energy. We use this correlation to sharpen predictions of the binding energy of neutron stars and the total neutrino energy in supernovae.  We find a minimum binding energy for a neutron star formed in a supernova of $\sim1.5\times 10^{53}$ ergs.  Should the neutrino energy in a supernova be significantly below this value, it would strongly suggest new unobserved particles are carrying away some of the supernova energy.  Alternatively, if the neutrino energy is observed above $\sim 6\times 10^{53}$ ergs, it would strongly imply the formation of a (perhaps surprisingly) massive neutron star.
\end{abstract}

\maketitle
\section{Introduction}

The next galactic core collapse supernova (SN) will be a watershed event. This SN may be an important gravitational wave source that could be the first multi-messenger event with gravitational wave, neutrino, and electromagnetic detections. The observation of about 20 neutrinos during SN1987A constrained the total neutrino energy to be $2-4\times 10^{53}$ ergs \cite{Hirata:1987,Bionta:1987}.  However, this event had both large statistical and systematic errors and only electron antineutrinos were likely observed.  Other neutrino flavors carried away significant energy and their unmeasured spectra and luminosities contribute substantially to systematic errors.

Modern neutrino detectors should observe many thousands of events from a SN at 10 kpc, dramatically increasing the statistics compared to SN1987A. Large neutrino detectors such as Super-Kamiokande \cite{Fukuda:2002uc} and Hyper-Kamiokande \cite{Abe:2018uyc} should provide excellent statistics on $\bar\nu_e$ while DUNE \cite{Goodman:2015gmv} will be a good $\nu_e$ detector.  To determine the total neutrino energy it is important to also measure muon and tauon neutrinos and antineutrinos.  This can be done via neutrino-nucleus elastic scattering \cite{PhysRevD.68.023005}.  Although the nuclear recoils in these detectors have low energies, the coherent cross section is very large, one is sensitive to all flavors of (active) neutrinos, and the bulk of the detector mass contributes.  As a result elastic scattering detectors can have yields of tens of events per {\it ton} for a SN at 10 kpc compared to 100s of events per {\it kilo-ton} for a conventional detector  \cite{PhysRevD.68.023005}.

A number of large dark matter detectors are now sensitive to a galactic SN via neutrino-nucleus elastic scattering.  Furthermore larger detectors, that are presently under construction or planned, should improve the statistics.  Detectors using Xe include Xenon1T \cite{Xenon1t,Aprile:2020tmw}, XenonnT, and DARWIN \cite{Darwin}. Large Ar detectors include Deap-3600 \cite{PhysRevD.100.022004}, Dark-Side, DarkSide-20k \cite{Dark-side-20k} and ARGO \cite{ARGO}.  A tens of tons dark matter detector with low threshold and good control of systematic errors can be a powerful supernova observatory and may be able to determine the total energy in active neutrinos to better than 10 percent \cite{PhysRevD.94.103009}.  In addition, RES-NOVA is a very promising idea for a Lead-based coherent scattering detector \cite{Pattavina:2020cqc}.


In this work, we focus on the total energy in (active) neutrinos radiated by a galactic SN.  This is a fundamental quantity that is insensitive to (active) neutrino oscillations and will be important for many analyses. The total neutrino energy is closely related to the gravitational binding energy, $E_b$, of the newly formed neutron star (NS).  If the total energy is observed to be low this could strongly suggest that new unobserved particles, such as sterile neutrinos or axions, carry away the remainder of the binding energy \cite{Particlebounds,PhysRevLett.60.1793}.  Alternatively, if the energy is large it may uniquely signal the formation of a (perhaps unexpectedly) massive neutron star.    

The binding energy of a neutron star has uncertainties from uncertainties in the equation of state (EOS). The EOS of dense matter governs the structure and behavior of nuclear matter in both macroscopic (e.g. neutron stars) and microscopic (e.g. atomic nuclei) systems. While the exact form of the EOS of dense matter is not fully understood, a great number of effective formalisms have been used to accurately predict the behavior of many macroscopic and microscopic nuclear systems. The EOS of nuclear matter has only been fully probed in laboratory experiments near saturation density, $\rho_0\approx 0.15$ fm$^{-3}$, and so one must look to astrophysical systems such as neutron stars to further probe the EOS. In the high density regime of a neutron star, the nuclear matter behaves at the zero temperature (``cold'') limit and can reach densities that are several times saturation density. 

Many astronomical radio \cite{demorest,2.14Msun}, X-ray \cite{Miller_2019} and gravitational wave \cite{Abbott:PRL2017, gw190425} observations have been able to provide constraints on neutron star observables and the high density EOS. Some NS observables closely related to the EOS are the mass (M), radius (R), moment of inertia (I), and dimensionless tidal polarizability ($\Lambda$). The \emph{tidal polarizability} (or deformability) is an intrinsic NS property highly sensitive to the stellar compactness $\xi=2GM/R$ \cite{Hinderer:2007mb,Hinderer:2009ca,Damour:2009vw,Postnikov:2010yn,Fattoyev:2012uu,Steiner:2014pda} that describes the tendency of a NS to develop a mass quadrupole as a response to the tidal field induced by its companion\,\cite{Damour:1991yw,Flanagan:2007ix}. The dimensionless tidal polarizability $\Lambda$ is defined as follows,
\begin{equation}
\Lambda = \frac{2}{3}k_{2} \Big(\frac{R}{GM}\Big)^{5}=\frac{64}{3}k_{2} \Big(\frac{R}{R_s}\Big)^{5}
\label{eq.Lambda}
\end{equation}
where $k_{2}$ is the second Love number\,\cite{Binnington:2009bb,Damour:2012yf}, and $M$ and $R$ are the neutron star mass and radius, respectively, and $R_s=2GM$. The limit on the deformability of a 1.4M$_\odot$ NS at 90\% confidence from GW170817 is $\Lambda_{1.4}=190_{-120}^{+390}$ \cite{LIGO_new}. This tidal deformability measurement rules out several previously viable EOSs \cite{Abbott:PRL2017}. Below, we show that this also improves our knowledge of the binding energy of neutron stars and increases the utility of measuring the total neutrino energy in a SN.

In this paper, we explore connections between $E_b$ and the tidal deformability $\Lambda$. Since both quantities are linked to the stellar compactness, we believe that constraints on either quantity from future observations can have meaningful constraints on the other.  Note that Sun {\it et al.} have explored the relationship between $E_b$ and compactness \cite{PhysRevD.102.023039}. In Sec. II, we detail the calculation of the gravitational binding energy of a NS and the dimensionless tidal deformability for various EOSs. Our selection of EOSs include both relativistic and non-relativistic models. We present our results including the behavior of $E_b$ on the deformability of a 1.4$\Msun$ NS ($\Lambda_{1.4}$) and on the mass of the NS in Sec. III. We discuss our results and implications for future supernova neutrino observations in Sec. IV and conclude in Sec. V. We use standard units where $c=\hbar=1$.

\begin{table}[htb]
\begin{tabular}{c|c|c|c|c}\hline\hline
EOS & $\Lambda_{1.4}$ & $E_b (1.174\Msun)$&$E_b (1.4\Msun)$ & $E_b (2\Msun)$\\
& &  [$10^{53}$ erg]&  [$10^{53}$ erg]&  [$10^{53}$ erg]\\\hline
APR \cite{Akmal:1998} & 252.7 & 1.82 & 2.74 & 6.51 \\
BSk20 \cite{Potekhin:2013}& 323.7 & 1.78 & 2.65 & 6.24 \\
BSk21 \cite{Potekhin:2013}& 516.1 & 1.68 & 2.48 & 5.65 \\
Big Apple \cite{big-apple:2020} & 713.0 & 1.62 & 2.38 & 5.26 \\
FPS \cite{Friedman:1981qw}& 177.4 & 1.86 & 2.82 & -- \\
FSUGold 2 \cite{Chen:2014sca}& 874.1 & 1.44 & 2.14 & 4.97 \\
FSUGold 2H \cite{Tolos:2017}& 782.6 & 1.60 & 2.35 & 5.21 \\
FSUGold 2R \cite{Tolos:2017}& 617.0 & 1.62 & 2.38 & 5.46 \\
FSUGarnet \cite{Chen:2014mza}& 642.7 & 1.61 & 2.38 & 5.43 \\
GNH3 \cite{Glendenning:1985}& 1094.6 & 1.31 & 1.98 & -- \\
IUFSU \cite{Fattoyev:2010mx}& 507.6 & 1.65 & 2.44 & -- \\
IUFSU2 \cite{Nandi:2019}& 562.7 & 1.64 & 2.42 & 5.55 \\
MPA1 \cite{Muther:1987}& 563.3 & 1.75 & 2.57 & 5.77 \\
NL3 \cite{Lalazissis:1996rd}& 1266.6 & 1.42 & 2.08 & 4.60 \\
NL3$\omega\rho$ (.209) \cite{Carriere:2002bx} & 952.2 & 1.54 & 2.25 & 4.94 \\
SLy \cite{Douchin:2000kx}& 320.1 & 1.72 & 2.59 & 6.28 \\
TAMUa \cite{Fattoyev:2013yaa}& 730.8 & 1.50 & 2.23 & 5.15 \\
TAMUb \cite{Fattoyev:2013yaa}& 979.6 & 1.40 & 2.07 & 4.79 \\
WFF2 \cite{Wiringa:1988tp}& 229.0 & 1.92 & 2.86 & 6.76 \\
\end{tabular}
\caption{Table of fiducial values for the EOSs used in the text. Missing values are when the EOS does not support a 2M$_\odot$ NS.}
\label{tab}
\end{table}
 
\section{Formalism}

\begin{figure*}[hbt]
    \centering
    \includegraphics[width=\textwidth]{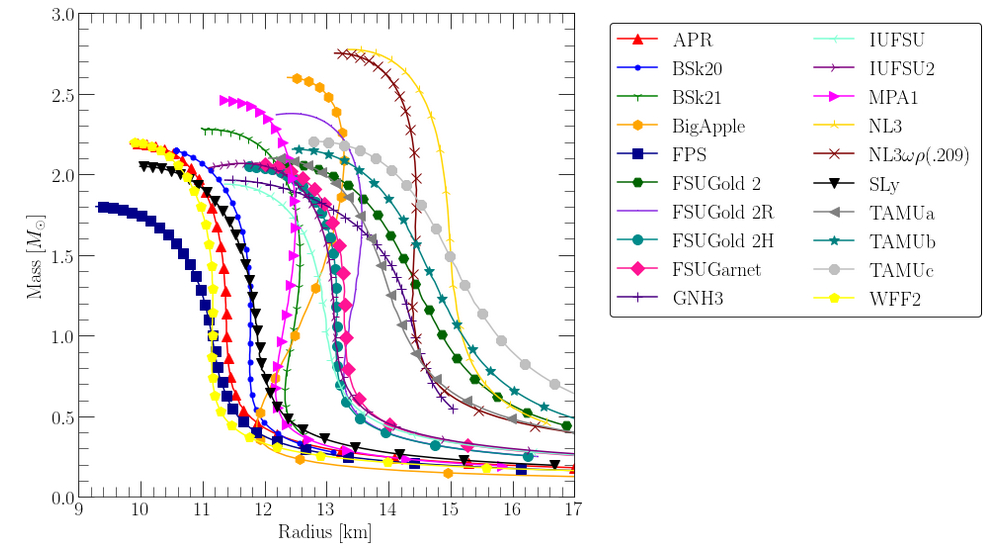}
    \caption{Mass-radius plot for the full set of EOS used in this study. The key for each EOS color and marker shape shall be used throughout the remainder of the figures in this paper.}
    \label{fig:mvr}
\end{figure*}

We calculate the binding energy of a NS in Sec. \ref{IIA} and deformability in Sec. \ref{IIB}. See Table \ref{tab} for a complete list of all EOSs used in this work. We chose this particular set of EOS as they predict a broad range of different properties of finite nuclei and NSs, in particular the maximum masses and the radii of 1.4$\Msun$ NSs explore a wide range of values, see Fig. \ref{fig:mvr}. This is by no means an exhaustive list of all possible NS EOS, however it does provide a range of very different predictions for both NS and finite nuclei properties.

For building our relativistic NS EOS, we adopt the Duflo-Zuker mass table for the outer crust defined up to a density of $4.3\times 10^{11}$ g cm$^{-3}$ \cite{Duflo:1995}. The inner crust is calculated by interpolating between the outer crust and liquid core with a cubic polynomial which is continuous in both the pressure and in the speed of sound \cite{piekarewicz:2019}. The core-crust transition density is given from the RPA formalism described in \cite{Carriere:2002bx}. Note that a number of these EOSs were downloaded from the web page of Feryal \"Ozel \cite{Ozel}.

\subsection{Neutron Star Binding Energy}
\label{IIA}

We define the gravitational mass of a NS to be the mass observed at infinity. This quantity is calculated by solving
    \begin{eqnarray}
    M=\int_0^R 4\pi r^2\rho(r)dr
    \label{eq.M}
    \end{eqnarray}
    in conjunction with the Tolman-Oppenheimer-Volkoff (TOV) equations to solve for the structure of a NS \cite{Opp39_PR55,Glendenning:2000}. Here, $\rho(r)$ is the energy density of matter.
    
    For a non-rotating NS, the volume element in the Schwarzschild metric takes the form
    \begin{eqnarray}
    dV=4\pi r^2 \Big [ 1-\frac{2GM(r)}{r}\Big ]^{-1/2}dr
    \end{eqnarray}
    thereby making the total number of baryons calculated from
    \begin{eqnarray}
    A=\int_0^R 4\pi r^2n(r)\Big[1-\frac{2GM(r)}{r}\Big ]^{-1/2}dr
    \label{eq.A}
    \end{eqnarray}
where $n(r)$ is the number density of baryons at a radius $r$. The baryonic mass of a NS can then be defined as the total mass of $A$ baryons, generally $M_b=m_\star A$ where $m_\star$ is the mass of a baryon (939 MeV) \cite{Lattimer:2006xb,Lattimer:2000nx,Weinberg:1972}. 
    
The binding energy of a NS is defined as the total amount of energy required to assemble $A$ baryons from infinity to form a stable star. This quantity can be measured via supernova neutrinos, which radiate approximately 99\% of the gravitational energy released from a core-collapse at the end of a massive star's life \cite{Prakash:1997,Lattimer:2000nx}. Since the collapsing object starts as a white-dwarf-like iron core, the total effective binding energy is calculated starting not from an ensemble of free baryons, but instead from iron nuclei,
    \begin{eqnarray}
    E_b = M-A\,\tilde m,
    \label{eq.ebe}
    \end{eqnarray}
where $\tilde m$ is the effective mass per baryon, taken to be the mass of $^{56}$Fe/56 $=930.412$ MeV \cite{Lattimer:2000nx}. We adopt a positive value for the binding energy and hereafter, references to ``binding energy'' will refer to the effective binding energy given by Eq.~\ref{eq.ebe}.
    
\subsection{Tidal Deformability Calculation}
\label{IIB}

We now detail the deformability calculation.  After one solves the TOV equations for the structure of a NS, we then calculate the second tidal Love number $k_2$ to solve for $\Lambda$. The second tidal Love number is calculated via
\begin{multline}
\label{love_num}
k_2 = \frac{1}{20}\xi^5(1-\xi)^2\Big[(2-y_R)+(y_R-1)\xi\Big]\times \\  \Big\{\Big[(6-3y_R)+\frac{3}{2}(5y_R-8)\xi\Big]\xi+\\
\frac{1}{2}\Big[(13-11y_R)+\frac{1}{2}(3y_R-2)\xi+\frac{1}{2}(1+y_R)\xi^2\Big]\xi^2+\\ 3\Big[(2-y_R)+(y_R-1)\xi\Big](1-\xi)^2\ln(1-\xi)\Big\}^{-1}\, .
\end{multline}
Here $\xi$ is the stellar compactness defined as before as $\xi=2GM/R$ and $y_R=y(R)$ is a dimensionless quantity which is calculated by solving the nonlinear differential equation,
\begin{equation}
    r\frac{dy}{dr}+y^2+F(r)y+r^2Q(r)=0
    \label{y_ode}
\end{equation}
with the initial condition $y(0)=2$. The functions $F(r)$ and $Q(r)$ are given below \cite{Hinderer:2007mb,piekarewicz:2019},
\begin{eqnarray}
&F(r) = \dfrac{r-4\pi Gr^3(\rho(r)-p(r))}{r-2GM(r)}\, , \\
&Q(r) = \dfrac{4\pi r}{r-2GM(r)}\Big[G\Big(5\rho(r)+9p(r)+\dfrac{\rho(r)+p(r)}{c_s^2(r)}\Big)\nonumber\\
&-\dfrac{6}{4\pi r^2}\Big]- 4\Big[\dfrac{G(M(r)+4\pi r^3p(r))}{r(r-2GM(r))}\Big]^2\, .
\end{eqnarray}
For a given EOS, we calculate the binding energy from Eqs. \ref{eq.M}, \ref{eq.A}, and \ref{eq.ebe} and the deformability from Eqs. \ref{eq.Lambda} and \ref{love_num} with $y_R$ from the solution to Eq. \ref{y_ode}.
\begin{figure*}[htb]
    \centering
    \includegraphics[width=\textwidth]{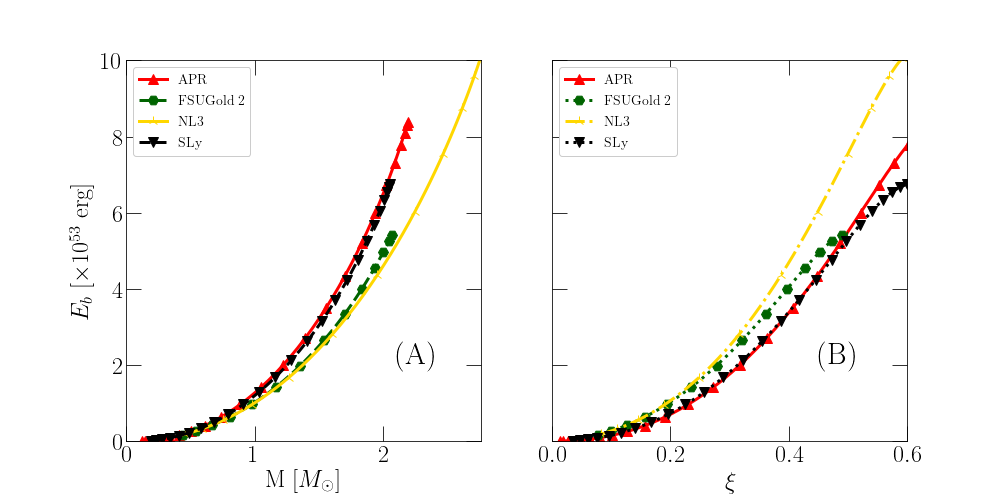}
    \caption{Gravitational binding energy versus stellar mass (A) and compactness $\xi=2GM/R$ (B) for four EOSs considered in the text. Fiducial values for all EOS can be found in Table \ref{tab}.}
\label{fig:eb_all}
\end{figure*}
\begin{figure}[htb]
    \centering
    \includegraphics[width=\columnwidth]{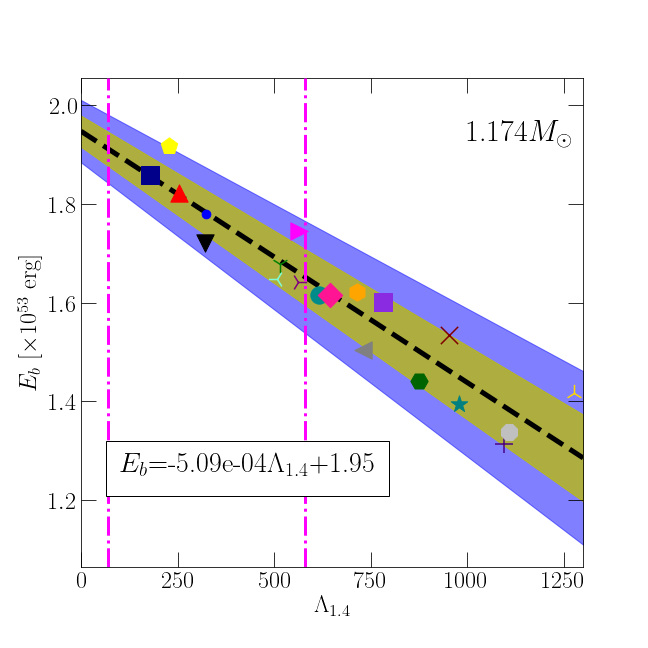}
\caption{Binding energy of a $1.174 \Msun$ NS versus $\Lambda_{1.4}$. The vertical dot-dashed lines show the upper and lower limits on $\Lambda_{1.4}$ from analysis of GW170817 \cite{Abbott:PRL2017}. The dashed line is the linear least-squares best fit with the yellow and blue regions marking the 1$\sigma$ and 2$\sigma$ errors on the fit respectively. The equation results from a linear least-squares fit, with $E_b$ in units of $10^{53}$ erg.}
    \label{fig:1.2}
\end{figure}
\begin{figure}[htb]
    \centering
    \includegraphics[width=\columnwidth]{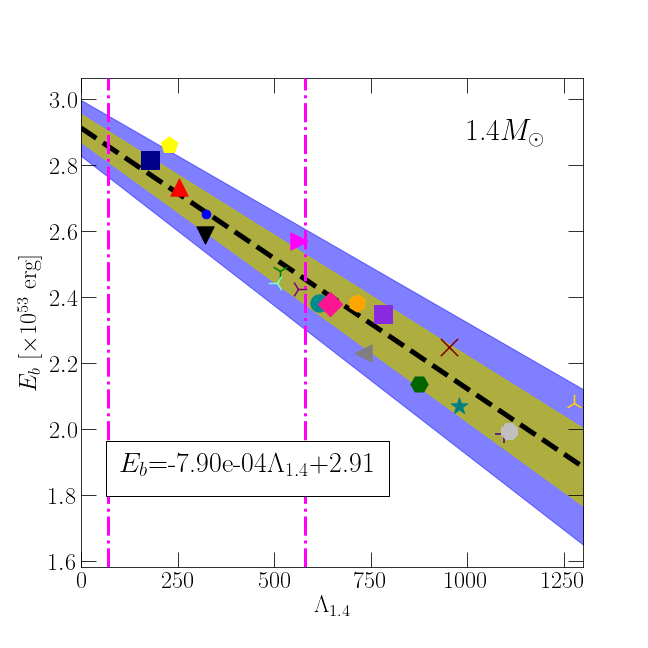}
\caption{Same as Fig. \ref{fig:1.2} but for 1.4$\Msun$ star.}
    \label{fig:1.4}
\end{figure}
\begin{figure}[htb]
    \centering
    \includegraphics[width=\columnwidth]{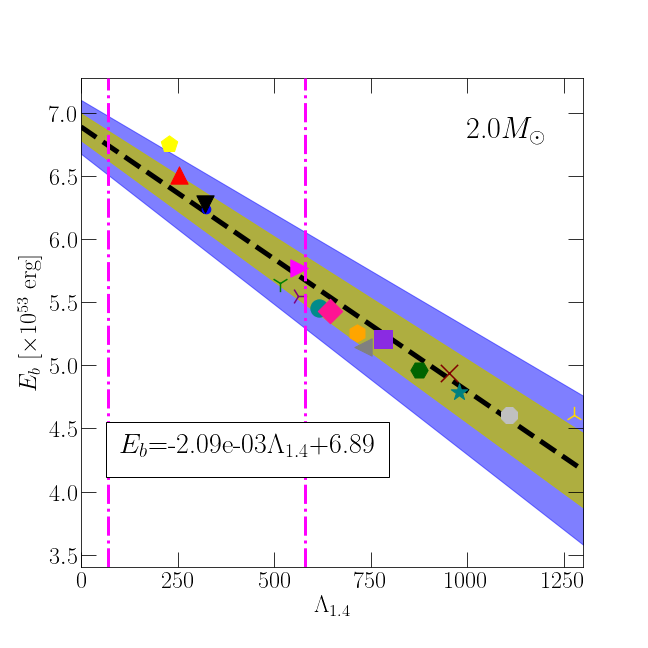}
\caption{Same as Fig. \ref{fig:1.2} but for 2.0$\Msun$ star.}
    \label{fig:2}
\end{figure}

\section{Results}  
\label{results}

We first show in Fig. \ref{fig:eb_all} the calculated binding energies for four of the twenty EOSs used in this study. We choose to plot $E_b$ against $\xi$ to illustrate its correlation with the stellar compactness. These EOSs were chosen so that we probe a variety of both relativistic (e.g. FSUGold 2 \cite{Chen:2014sca} and NL3 \cite{Lalazissis:1996rd}) and non-relativistic (e.g. SLy \cite{Douchin:2000kx} and APR \cite{Akmal:1998}) formalisms. All EOSs along with fiducial values for the deformability and binding energies can be found in Table \ref{tab}.

We then plot the behavior of $E_b$ and $\Lambda_{1.4}$ for fixed values of $M$ in Figs. \ref{fig:1.2}-\ref{fig:2}. These figures reveal a linear correlation between $E_b$ and $\Lambda_{1.4}$ for a fixed value of $M$. If one were to fix $\Lambda_{1.4}$, this may provide a simple way to estimate the binding energy or gravitational mass should one observe either quantity more stringently.  A simple least-squares fit $E_b\approx a_1 \Lambda_{1.4}+a_2$ yields the parameters $a_1$ and $a_2$ along with the 1$\sigma$ and 2$\sigma$ error bands as shown in Figs. \ref{fig:1.2}-\ref{fig:2}. The $1\sigma$ values are calculated from the estimated covariance matrix generated using the SciPy  \cite{2020SciPy-NMeth} package curve\_fit() \cite{curve_fit}, which uses the Levenberg-Marquardt minimization routine to obtain an optimal parameter set. The 2$\sigma$ values are then just twice the 1$\sigma$ values.

\begin{figure}[htb]
    \centering
    \includegraphics[width=.9\columnwidth]{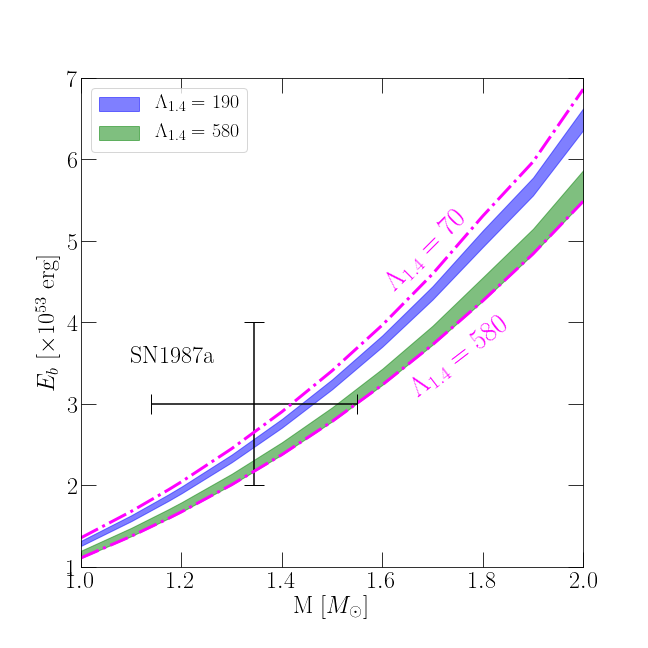}
\caption{Binding energy as a function of remnant mass at various fixed deformabilities. The blue and green bands show the 1$\sigma$ confidence intervals based on the linear fits described in Sec. \ref{results} for $\Lambda_{1.4}=190$ and  $580$ respectively. The black error bars represent the constraints from SN1987a \cite{Hirata:1987,Bionta:1987}. The dash-dot pink curves represent the bounds of allowed $E_b$ for deformabilities in the range $\Lambda_{1.4}=[70,580]$.}
    \label{fig:eb_m580}
\end{figure}

These error regions can then be used to display the dependency of $E_b$ on $M$ and provide a way to estimate either if one fixes $\Lambda_{1.4}$. As our knowledge of $\Lambda_{1.4}$ improves, a measurement of $E_b$ more sharply constrains $M$.  To illustrate this, we use the linear fits described above to calculate a 1$\sigma$ band of values for $E_b$ at different values of $M$ for a few characteristic values of $\Lambda_{1.4}$ in Fig.~\ref{fig:eb_m580}. We observe a strongly constrained band of $E_b$ values at each $M$ for each of the characteristic $\Lambda_{1.4}$ values. For example, should future gravitational wave observations constrain $\Lambda_{1.4}$ to be very close to the GW170817 central value $\approx190$, this band becomes further constrained to higher values of $E_b$ with less uncertainty.  For comparison, we also show the estimates of $E_b$ and the remnant mass from SN1987a \cite{Hirata:1987,Bionta:1987}.

We conclude these results by presenting constraints on the binding energy of low mass NSs. We show the calculated $E_b$ values for a few different mass NSs, fixing $\Lambda_{1.4}=580$ in Table \ref{tab:EbM1} and $\Lambda_{1.4}=190$ in Table \ref{tab:EbM2}. In particular, we show the $E_b$ for the lowest observed NS mass of $1.174\pm 0.004\Msun$ \cite{Martinez_2015}. We find that based on our linear fitting procedure above, the lowest mass neutron star must have a binding energy $\gtrapprox 1.5\times 10^{53}$ erg.

\begin{table}[htb]
\begin{tabular}{c|c|c}\hline\hline
Mass & $E_b(\Lambda_{1.4}=580)$ & $2\sigma$ Range \\
$[\Msun]$ & [$10^{53}$ erg] & [$10^{53}$ erg]\\\hline
1.0 & 1.16 $\pm$ 0.044 & [1.067, 1.244] \\
1.174 & 1.65 $\pm$ 0.057 & [1.539, 1.766] \\
1.4 & 2.45 $\pm$ 0.076 & [2.302, 2.606] \\
1.8 & 4.41 $\pm$ 0.144 & [4.120, 4.697] \\
2.0 & 5.68 $\pm$ 0.191 & [5.294, 6.057] \\
\end{tabular}
\caption{Estimated binding energies for different remnant masses assuming a deformability of $\Lambda_{1.4}=580$. The middle column shows the 1$\sigma$ error while the right column shows the 2$\sigma$ range of binding energies for the given remnant mass.}
\label{tab:EbM1}
\end{table}

\begin{table}[htb]
\begin{tabular}{c|c|c}\hline\hline
Mass & $E_b(\Lambda_{1.4}=190)$ & $2\sigma$ Range \\
$[\Msun]$ & [$10^{53}$ erg] & [$10^{53}$ erg]\\\hline
1.0 & 1.29 $\pm$ 0.031 & [1.228, 1.352] \\
1.174 & 1.85 $\pm$ 0.040 & [1.771, 1.930] \\
1.4 & 2.76 $\pm$ 0.053 & [2.656, 2.869] \\
1.8 & 5.03 $\pm$ 0.101 & [4.823, 5.228] \\
2.0 & 6.49 $\pm$ 0.134 & [6.223, 6.760] \\
\end{tabular}
\caption{Same as Table \ref{tab:EbM1} but for $\Lambda_{1.4}=190$.}
\label{tab:EbM2}
\end{table}

\section{Discussion and Implications}
We will now discuss the implications of our findings. Firstly, the gravitational binding energy is correlated with the deformability $\Lambda_{1.4}$. This correlation can be used to calculate an expected range of values for $E_b$ at a given $M$. Different values for $\Lambda_{1.4}$ predict different curves for the behavior of $E_b$ and this can be used to place constraints on either $\Lambda_{1.4}$ or $E_b$ should the other become better constrained. As our knowledge of $\Lambda_{1.4}$ grows from additional gravitational wave observations, $E_b$ will become better constrained.  This increases the utility of measuring the total energy in active neutrinos  during the next galactic supernova.  

Since the range of $E_b$ predicted is fairly well constrained at each $M$, this allows important tests for new physics. For instance, the existence of new exotic particles beyond the standard model may affect the energetics of the supernova.  The lowest, well measured, neutron star mass is $1.174\pm 0.004$M$_\odot$ \cite{Martinez_2015}. The existence of such a low mass NS has provided a challenge for supernova simulations as it is difficult to achieve such a low-mass remnant \cite{suwa:2018}. It may also be difficult to produce a much lighter NS in a SN because the star's core must reach the Chandrasekhar mass, $\approx 1.4\Msun$, in order to collapse, and the SN shock may only remove a small fraction of this mass.  This leaves behind a NS not significantly lighter than 1.174M$_\odot$.   

We find an important {\it lower bound} on $E_b$ for a NS.  Constraints on $\Lambda_{1.4}<580$ now rule out stiff equations of state that predict low $E_b$.  For $\Lambda_{1.4}=580$, we find a $2\sigma$ lower bound for a $1.174$M$_\odot$ star of $E_b>1.52\times 10^{53}$ ergs.  This provides a {\it lower bound on the energy radiated in active neutrinos} during a SN.  If the observed energy is significantly less than this bound, it is very likely some of the NS binding energy went into new unobserved particles such as sterile neutrinos, axions, or other light dark matter constituents \cite{Particlebounds,PhysRevLett.60.1793}.  This emphasizes that the total energy in neutrinos during a SN is a fundamental quantity that is very important to measure well.

The gravitational wave event GW190814 involved a 2.6M$_\odot$ compact object that could be the most massive neutron star observed \cite{Abbott_2020}, although subsequent analyses disfavor this hypothesis \cite{big-apple:2020,Tews:2020,Lim:2020,Dexheimer:2020,Roupas:2020}. Nevertheless, a supernova birthing such a massive NS would radiate an extremely large energy in neutrinos. We can see from Fig. \ref{fig:eb_all} that most EOS predict a NS maximum mass of $\sim$2 $\Msun$ with a binding energy of $\sim$5-7 $\times 10^{53}$ erg. However, two EOS used here predict a maximum mass $\geq2.6\Msun$ with $E_b$ in excess of $10^{54}$ erg. If the energy in neutrinos from a supernova significantly exceeds $\sim 6\times 10^{53}$ erg, this would almost uniquely imply the formation of a very massive NS.

\medskip

\section{Conclusion}

We have used several different equations of state to predict that the binding energy of a neutron star $E_b$ is correlated with the tidal deformability. Using our correlation and the LIGO gravitational wave constraint on the deformability of a $1.4M_\odot$ NS of $\Lambda_{1.4}=190_{-120}^{+390}$, we determine a range of binding energies expected for a given neutron star mass. This range will be further constrained as $\Lambda_{1.4}$ becomes better determined with future gravitational wave observations.  The total energy radiated in neutrinos during a SN is closely related to $E_b$.  This reduction in the uncertainty in $E_b$ increases the utility and importance of accurately measuring the total energy in neutrinos from the next SN.  If the total energy is significantly less than $\sim1.5\times 10^{53}$ ergs it may strongly suggest that new particles are carrying away some of the SN energy.  Alternatively if the energy is unexpectedly large, it may uniquely imply the formation of a surprisingly massive NS.

\begin{acknowledgments}
We would like to thank the anonymous referee for their suggested improvements to this work. We would also like to thank Alex Deibel for providing the APR equation of state.
This material is based upon work supported by the U.S. Department of Energy Office of Science, Office of Nuclear Physics under Awards DE-FG02-87ER40365 (Indiana University) and Number DE-SC0008808 (NUCLEI SciDAC Collaboration). 

\end{acknowledgments}


\begin{thebibliography}{67}
\expandafter\ifx\csname natexlab\endcsname\relax\def\natexlab#1{#1}\fi
\expandafter\ifx\csname bibnamefont\endcsname\relax
  \def\bibnamefont#1{#1}\fi
\expandafter\ifx\csname bibfnamefont\endcsname\relax
  \def\bibfnamefont#1{#1}\fi
\expandafter\ifx\csname citenamefont\endcsname\relax
  \def\citenamefont#1{#1}\fi
\expandafter\ifx\csname url\endcsname\relax
  \def\url#1{\texttt{#1}}\fi
\expandafter\ifx\csname urlprefix\endcsname\relax\def\urlprefix{URL }\fi
\providecommand{\bibinfo}[2]{#2}
\providecommand{\eprint}[2][]{\url{#2}}

\bibitem[{\citenamefont{{Hirata} et~al.}(1987)\citenamefont{{Hirata}, {Kajita},
  {Koshiba}, {Nakahata}, {Oyama}, {Sato}, {Suzuki}, {Takita}, {Totsuka},
  {Kifune} et~al.}}]{Hirata:1987}
\bibinfo{author}{\bibfnamefont{K.}~\bibnamefont{{Hirata}}},
  \bibinfo{author}{\bibfnamefont{T.}~\bibnamefont{{Kajita}}},
  \bibinfo{author}{\bibfnamefont{M.}~\bibnamefont{{Koshiba}}},
  \bibinfo{author}{\bibfnamefont{M.}~\bibnamefont{{Nakahata}}},
  \bibinfo{author}{\bibfnamefont{Y.}~\bibnamefont{{Oyama}}},
  \bibinfo{author}{\bibfnamefont{N.}~\bibnamefont{{Sato}}},
  \bibinfo{author}{\bibfnamefont{A.}~\bibnamefont{{Suzuki}}},
  \bibinfo{author}{\bibfnamefont{M.}~\bibnamefont{{Takita}}},
  \bibinfo{author}{\bibfnamefont{Y.}~\bibnamefont{{Totsuka}}},
  \bibinfo{author}{\bibfnamefont{T.}~\bibnamefont{{Kifune}}},
  \bibnamefont{et~al.}, \bibinfo{journal}{\prl} \textbf{\bibinfo{volume}{58}},
  \bibinfo{pages}{1490} (\bibinfo{year}{1987}).

\bibitem[{\citenamefont{{Bionta} et~al.}(1987)\citenamefont{{Bionta},
  {Blewitt}, {Bratton}, {Casper}, {Ciocio}, {Claus}, {Cortez}, {Crouch}, {Dye},
  {Errede} et~al.}}]{Bionta:1987}
\bibinfo{author}{\bibfnamefont{R.~M.} \bibnamefont{{Bionta}}},
  \bibinfo{author}{\bibfnamefont{G.}~\bibnamefont{{Blewitt}}},
  \bibinfo{author}{\bibfnamefont{C.~B.} \bibnamefont{{Bratton}}},
  \bibinfo{author}{\bibfnamefont{D.}~\bibnamefont{{Casper}}},
  \bibinfo{author}{\bibfnamefont{A.}~\bibnamefont{{Ciocio}}},
  \bibinfo{author}{\bibfnamefont{R.}~\bibnamefont{{Claus}}},
  \bibinfo{author}{\bibfnamefont{B.}~\bibnamefont{{Cortez}}},
  \bibinfo{author}{\bibfnamefont{M.}~\bibnamefont{{Crouch}}},
  \bibinfo{author}{\bibfnamefont{S.~T.} \bibnamefont{{Dye}}},
  \bibinfo{author}{\bibfnamefont{S.}~\bibnamefont{{Errede}}},
  \bibnamefont{et~al.}, \bibinfo{journal}{\prl} \textbf{\bibinfo{volume}{58}},
  \bibinfo{pages}{1494} (\bibinfo{year}{1987}).

\bibitem[{\citenamefont{Fukuda et~al.}(2003)}]{Fukuda:2002uc}
\bibinfo{author}{\bibfnamefont{Y.}~\bibnamefont{Fukuda}} \bibnamefont{et~al.}
  (\bibinfo{collaboration}{Super-Kamiokande}), \bibinfo{journal}{Nucl. Instrum.
  Meth. A} \textbf{\bibinfo{volume}{501}}, \bibinfo{pages}{418}
  (\bibinfo{year}{2003}).

\bibitem[{\citenamefont{Abe et~al.}(2018)}]{Abe:2018uyc}
\bibinfo{author}{\bibfnamefont{K.}~\bibnamefont{Abe}} \bibnamefont{et~al.}
  (\bibinfo{collaboration}{Hyper-Kamiokande}) (\bibinfo{year}{2018}),
  \eprint{1805.04163}.

\bibitem[{\citenamefont{Goodman}(2015)}]{Goodman:2015gmv}
\bibinfo{author}{\bibfnamefont{M.}~\bibnamefont{Goodman}},
  \bibinfo{journal}{Adv. High Energy Phys.} \textbf{\bibinfo{volume}{2015}},
  \bibinfo{pages}{256351} (\bibinfo{year}{2015}).

\bibitem[{\citenamefont{Horowitz et~al.}(2003)\citenamefont{Horowitz, Coakley,
  and McKinsey}}]{PhysRevD.68.023005}
\bibinfo{author}{\bibfnamefont{C.~J.} \bibnamefont{Horowitz}},
  \bibinfo{author}{\bibfnamefont{K.~J.} \bibnamefont{Coakley}},
  \bibnamefont{and} \bibinfo{author}{\bibfnamefont{D.~N.}
  \bibnamefont{McKinsey}}, \bibinfo{journal}{Phys. Rev. D}
  \textbf{\bibinfo{volume}{68}}, \bibinfo{pages}{023005}
  (\bibinfo{year}{2003}),
  \urlprefix\url{https://link.aps.org/doi/10.1103/PhysRevD.68.023005}.

\bibitem[{\citenamefont{Aprile et~al.}(2016)\citenamefont{Aprile, Aalbers,
  Agostini, Alfonsi, Amaro, Anthony, Arazi, Arneodo, Balan, Barrow
  et~al.}}]{Xenon1t}
\bibinfo{author}{\bibfnamefont{E.}~\bibnamefont{Aprile}},
  \bibinfo{author}{\bibfnamefont{J.}~\bibnamefont{Aalbers}},
  \bibinfo{author}{\bibfnamefont{F.}~\bibnamefont{Agostini}},
  \bibinfo{author}{\bibfnamefont{M.}~\bibnamefont{Alfonsi}},
  \bibinfo{author}{\bibfnamefont{F.~D.} \bibnamefont{Amaro}},
  \bibinfo{author}{\bibfnamefont{M.}~\bibnamefont{Anthony}},
  \bibinfo{author}{\bibfnamefont{L.}~\bibnamefont{Arazi}},
  \bibinfo{author}{\bibfnamefont{F.}~\bibnamefont{Arneodo}},
  \bibinfo{author}{\bibfnamefont{C.}~\bibnamefont{Balan}},
  \bibinfo{author}{\bibfnamefont{P.}~\bibnamefont{Barrow}},
  \bibnamefont{et~al.}, \bibinfo{journal}{Journal of Cosmology and
  Astroparticle Physics} \textbf{\bibinfo{volume}{2016}}, \bibinfo{pages}{027}
  (\bibinfo{year}{2016}).

\bibitem[{\citenamefont{Aprile et~al.}(2020)}]{Aprile:2020tmw}
\bibinfo{author}{\bibfnamefont{E.}~\bibnamefont{Aprile}} \bibnamefont{et~al.}
  (\bibinfo{collaboration}{XENON}) (\bibinfo{year}{2020}), \eprint{2006.09721}.

\bibitem[{\citenamefont{Aalbers et~al.}(2016)\citenamefont{Aalbers, Agostini,
  Alfonsi, Amaro, Amsler, Aprile, Arazi, Arneodo, Barrow, Baudis
  et~al.}}]{Darwin}
\bibinfo{author}{\bibfnamefont{J.}~\bibnamefont{Aalbers}},
  \bibinfo{author}{\bibfnamefont{F.}~\bibnamefont{Agostini}},
  \bibinfo{author}{\bibfnamefont{M.}~\bibnamefont{Alfonsi}},
  \bibinfo{author}{\bibfnamefont{F.}~\bibnamefont{Amaro}},
  \bibinfo{author}{\bibfnamefont{C.}~\bibnamefont{Amsler}},
  \bibinfo{author}{\bibfnamefont{E.}~\bibnamefont{Aprile}},
  \bibinfo{author}{\bibfnamefont{L.}~\bibnamefont{Arazi}},
  \bibinfo{author}{\bibfnamefont{F.}~\bibnamefont{Arneodo}},
  \bibinfo{author}{\bibfnamefont{P.}~\bibnamefont{Barrow}},
  \bibinfo{author}{\bibfnamefont{L.}~\bibnamefont{Baudis}},
  \bibnamefont{et~al.}, \bibinfo{journal}{Journal of Cosmology and
  Astroparticle Physics} \textbf{\bibinfo{volume}{2016}}, \bibinfo{pages}{017}
  (\bibinfo{year}{2016}).

\bibitem[{\citenamefont{Ajaj et~al.}(2019)\citenamefont{Ajaj, Amaudruz, Araujo,
  Baldwin, Batygov, Beltran, Bina, Bonatt, Boulay, Broerman
  et~al.}}]{PhysRevD.100.022004}
\bibinfo{author}{\bibfnamefont{R.}~\bibnamefont{Ajaj}},
  \bibinfo{author}{\bibfnamefont{P.-A.} \bibnamefont{Amaudruz}},
  \bibinfo{author}{\bibfnamefont{G.~R.} \bibnamefont{Araujo}},
  \bibinfo{author}{\bibfnamefont{M.}~\bibnamefont{Baldwin}},
  \bibinfo{author}{\bibfnamefont{M.}~\bibnamefont{Batygov}},
  \bibinfo{author}{\bibfnamefont{B.}~\bibnamefont{Beltran}},
  \bibinfo{author}{\bibfnamefont{C.~E.} \bibnamefont{Bina}},
  \bibinfo{author}{\bibfnamefont{J.}~\bibnamefont{Bonatt}},
  \bibinfo{author}{\bibfnamefont{M.~G.} \bibnamefont{Boulay}},
  \bibinfo{author}{\bibfnamefont{B.}~\bibnamefont{Broerman}},
  \bibnamefont{et~al.} (\bibinfo{collaboration}{DEAP Collaboration}),
  \bibinfo{journal}{Phys. Rev. D} \textbf{\bibinfo{volume}{100}},
  \bibinfo{pages}{022004} (\bibinfo{year}{2019}),
  \urlprefix\url{https://link.aps.org/doi/10.1103/PhysRevD.100.022004}.

\bibitem[{\citenamefont{Aalseth et~al.}(2018)\citenamefont{Aalseth, Acerbi,
  Agnes, Albuquerque, Alexander, Alici, Alton, Antonioli, Arcelli, Ardito
  et~al.}}]{Dark-side-20k}
\bibinfo{author}{\bibfnamefont{C.~E.} \bibnamefont{Aalseth}},
  \bibinfo{author}{\bibfnamefont{F.}~\bibnamefont{Acerbi}},
  \bibinfo{author}{\bibfnamefont{P.}~\bibnamefont{Agnes}},
  \bibinfo{author}{\bibfnamefont{I.~F.~M.} \bibnamefont{Albuquerque}},
  \bibinfo{author}{\bibfnamefont{T.}~\bibnamefont{Alexander}},
  \bibinfo{author}{\bibfnamefont{A.}~\bibnamefont{Alici}},
  \bibinfo{author}{\bibfnamefont{A.~K.} \bibnamefont{Alton}},
  \bibinfo{author}{\bibfnamefont{P.}~\bibnamefont{Antonioli}},
  \bibinfo{author}{\bibfnamefont{S.}~\bibnamefont{Arcelli}},
  \bibinfo{author}{\bibfnamefont{R.}~\bibnamefont{Ardito}},
  \bibnamefont{et~al.}, \bibinfo{journal}{The European Physical Journal Plus}
  \textbf{\bibinfo{volume}{133}}, \bibinfo{pages}{131} (\bibinfo{year}{2018}),
  \urlprefix\url{https://doi.org/10.1140/epjp/i2018-11973-4}.

\bibitem[{\citenamefont{Gonzalez-Garcia
  et~al.}(2018)\citenamefont{Gonzalez-Garcia, Maltoni, Perez-Gonzalez, and
  Funchal}}]{ARGO}
\bibinfo{author}{\bibfnamefont{M.~C.} \bibnamefont{Gonzalez-Garcia}},
  \bibinfo{author}{\bibfnamefont{M.}~\bibnamefont{Maltoni}},
  \bibinfo{author}{\bibfnamefont{Y.~F.} \bibnamefont{Perez-Gonzalez}},
  \bibnamefont{and} \bibinfo{author}{\bibfnamefont{R.~Z.}
  \bibnamefont{Funchal}}, \bibinfo{journal}{Journal of High Energy Physics}
  \textbf{\bibinfo{volume}{2018}}, \bibinfo{pages}{19} (\bibinfo{year}{2018}),
  \urlprefix\url{https://doi.org/10.1007/JHEP07(2018)019}.

\bibitem[{\citenamefont{Lang et~al.}(2016)\citenamefont{Lang, McCabe, Reichard,
  Selvi, and Tamborra}}]{PhysRevD.94.103009}
\bibinfo{author}{\bibfnamefont{R.~F.} \bibnamefont{Lang}},
  \bibinfo{author}{\bibfnamefont{C.}~\bibnamefont{McCabe}},
  \bibinfo{author}{\bibfnamefont{S.}~\bibnamefont{Reichard}},
  \bibinfo{author}{\bibfnamefont{M.}~\bibnamefont{Selvi}}, \bibnamefont{and}
  \bibinfo{author}{\bibfnamefont{I.}~\bibnamefont{Tamborra}},
  \bibinfo{journal}{Phys. Rev. D} \textbf{\bibinfo{volume}{94}},
  \bibinfo{pages}{103009} (\bibinfo{year}{2016}),
  \urlprefix\url{https://link.aps.org/doi/10.1103/PhysRevD.94.103009}.

\bibitem[{\citenamefont{Pattavina et~al.}(2020)\citenamefont{Pattavina,
  Iachellini, and Tamborra}}]{Pattavina:2020cqc}
\bibinfo{author}{\bibfnamefont{L.}~\bibnamefont{Pattavina}},
  \bibinfo{author}{\bibfnamefont{N.~F.} \bibnamefont{Iachellini}},
  \bibnamefont{and} \bibinfo{author}{\bibfnamefont{I.}~\bibnamefont{Tamborra}}
  (\bibinfo{year}{2020}), \eprint{2004.06936}.

\bibitem[{\citenamefont{Chang et~al.}(2018)\citenamefont{Chang, Essig, and
  McDermott}}]{Particlebounds}
\bibinfo{author}{\bibfnamefont{J.~H.} \bibnamefont{Chang}},
  \bibinfo{author}{\bibfnamefont{R.}~\bibnamefont{Essig}}, \bibnamefont{and}
  \bibinfo{author}{\bibfnamefont{S.~D.} \bibnamefont{McDermott}},
  \bibinfo{journal}{Journal of High Energy Physics}
  \textbf{\bibinfo{volume}{2018}}, \bibinfo{pages}{51} (\bibinfo{year}{2018}),
  \urlprefix\url{https://doi.org/10.1007/JHEP09(2018)051}.

\bibitem[{\citenamefont{Raffelt and Seckel}(1988)}]{PhysRevLett.60.1793}
\bibinfo{author}{\bibfnamefont{G.}~\bibnamefont{Raffelt}} \bibnamefont{and}
  \bibinfo{author}{\bibfnamefont{D.}~\bibnamefont{Seckel}},
  \bibinfo{journal}{Phys. Rev. Lett.} \textbf{\bibinfo{volume}{60}},
  \bibinfo{pages}{1793} (\bibinfo{year}{1988}),
  \urlprefix\url{https://link.aps.org/doi/10.1103/PhysRevLett.60.1793}.

\bibitem[{\citenamefont{Demorest et~al.}(2010)\citenamefont{Demorest, Pennucci,
  Ransom, Roberts, and Hessels}}]{demorest}
\bibinfo{author}{\bibfnamefont{P.~B.} \bibnamefont{Demorest}},
  \bibinfo{author}{\bibfnamefont{T.}~\bibnamefont{Pennucci}},
  \bibinfo{author}{\bibfnamefont{S.~M.} \bibnamefont{Ransom}},
  \bibinfo{author}{\bibfnamefont{M.~S.~E.} \bibnamefont{Roberts}},
  \bibnamefont{and} \bibinfo{author}{\bibfnamefont{J.~W.~T.}
  \bibnamefont{Hessels}}, \bibinfo{journal}{Nature}
  \textbf{\bibinfo{volume}{467}}, \bibinfo{pages}{1081 EP }
  (\bibinfo{year}{2010}), \urlprefix\url{https://doi.org/10.1038/nature09466}.

\bibitem[{\citenamefont{Cromartie et~al.}(2019)\citenamefont{Cromartie,
  Fonseca, Ransom, Demorest, Arzoumanian, Blumer, Brook, DeCesar, Dolch, Ellis
  et~al.}}]{2.14Msun}
\bibinfo{author}{\bibfnamefont{H.~T.} \bibnamefont{Cromartie}},
  \bibinfo{author}{\bibfnamefont{E.}~\bibnamefont{Fonseca}},
  \bibinfo{author}{\bibfnamefont{S.~M.} \bibnamefont{Ransom}},
  \bibinfo{author}{\bibfnamefont{P.~B.} \bibnamefont{Demorest}},
  \bibinfo{author}{\bibfnamefont{Z.}~\bibnamefont{Arzoumanian}},
  \bibinfo{author}{\bibfnamefont{H.}~\bibnamefont{Blumer}},
  \bibinfo{author}{\bibfnamefont{P.~R.} \bibnamefont{Brook}},
  \bibinfo{author}{\bibfnamefont{M.~E.} \bibnamefont{DeCesar}},
  \bibinfo{author}{\bibfnamefont{T.}~\bibnamefont{Dolch}},
  \bibinfo{author}{\bibfnamefont{J.~A.} \bibnamefont{Ellis}},
  \bibnamefont{et~al.}, \bibinfo{journal}{Nature Astronomy}
  (\bibinfo{year}{2019}),
  \urlprefix\url{https://doi.org/10.1038/s41550-019-0880-2}.

\bibitem[{\citenamefont{Miller et~al.}(2019)\citenamefont{Miller, Lamb,
  Dittmann, Bogdanov, Arzoumanian, Gendreau, Guillot, Harding, Ho, Lattimer
  et~al.}}]{Miller_2019}
\bibinfo{author}{\bibfnamefont{M.~C.} \bibnamefont{Miller}},
  \bibinfo{author}{\bibfnamefont{F.~K.} \bibnamefont{Lamb}},
  \bibinfo{author}{\bibfnamefont{A.~J.} \bibnamefont{Dittmann}},
  \bibinfo{author}{\bibfnamefont{S.}~\bibnamefont{Bogdanov}},
  \bibinfo{author}{\bibfnamefont{Z.}~\bibnamefont{Arzoumanian}},
  \bibinfo{author}{\bibfnamefont{K.~C.} \bibnamefont{Gendreau}},
  \bibinfo{author}{\bibfnamefont{S.}~\bibnamefont{Guillot}},
  \bibinfo{author}{\bibfnamefont{A.~K.} \bibnamefont{Harding}},
  \bibinfo{author}{\bibfnamefont{W.~C.~G.} \bibnamefont{Ho}},
  \bibinfo{author}{\bibfnamefont{J.~M.} \bibnamefont{Lattimer}},
  \bibnamefont{et~al.}, \bibinfo{journal}{The Astrophysical Journal}
  \textbf{\bibinfo{volume}{887}}, \bibinfo{pages}{L24} (\bibinfo{year}{2019}),
  \urlprefix\url{https://doi.org/10.3847%2F2041-8213%2Fab50c5}.

\bibitem[{\citenamefont{Abbott et~al.}(2017)}]{Abbott:PRL2017}
\bibinfo{author}{\bibfnamefont{B.~P.} \bibnamefont{Abbott}}
  \bibnamefont{et~al.} (\bibinfo{collaboration}{Virgo, LIGO Scientific}),
  \bibinfo{journal}{Phys. Rev. Lett.} \textbf{\bibinfo{volume}{119}},
  \bibinfo{pages}{161101} (\bibinfo{year}{2017}).

\bibitem[{\citenamefont{{Abbott} et~al.}(2020)\citenamefont{{Abbott}, {Abbott},
  {Abbott}, {Abraham}, {Acernese}, {Ackley}, {Adams}, {Adhikari}, {Adya},
  {Affeldt} et~al.}}]{gw190425}
\bibinfo{author}{\bibfnamefont{B.~P.} \bibnamefont{{Abbott}}},
  \bibinfo{author}{\bibfnamefont{R.}~\bibnamefont{{Abbott}}},
  \bibinfo{author}{\bibfnamefont{T.~D.} \bibnamefont{{Abbott}}},
  \bibinfo{author}{\bibfnamefont{S.}~\bibnamefont{{Abraham}}},
  \bibinfo{author}{\bibfnamefont{F.}~\bibnamefont{{Acernese}}},
  \bibinfo{author}{\bibfnamefont{K.}~\bibnamefont{{Ackley}}},
  \bibinfo{author}{\bibfnamefont{C.}~\bibnamefont{{Adams}}},
  \bibinfo{author}{\bibfnamefont{R.~X.} \bibnamefont{{Adhikari}}},
  \bibinfo{author}{\bibfnamefont{V.~B.} \bibnamefont{{Adya}}},
  \bibinfo{author}{\bibfnamefont{C.}~\bibnamefont{{Affeldt}}},
  \bibnamefont{et~al.}, \bibinfo{journal}{\apjl}
  \textbf{\bibinfo{volume}{892}}, \bibinfo{eid}{L3} (\bibinfo{year}{2020}),
  \eprint{2001.01761}.

\bibitem[{\citenamefont{Hinderer}(2008)}]{Hinderer:2007mb}
\bibinfo{author}{\bibfnamefont{T.}~\bibnamefont{Hinderer}},
  \bibinfo{journal}{Astrophys. J.} \textbf{\bibinfo{volume}{677}},
  \bibinfo{pages}{1216} (\bibinfo{year}{2008}).

\bibitem[{\citenamefont{Hinderer et~al.}(2010)\citenamefont{Hinderer, Lackey,
  Lang, and Read}}]{Hinderer:2009ca}
\bibinfo{author}{\bibfnamefont{T.}~\bibnamefont{Hinderer}},
  \bibinfo{author}{\bibfnamefont{B.~D.} \bibnamefont{Lackey}},
  \bibinfo{author}{\bibfnamefont{R.~N.} \bibnamefont{Lang}}, \bibnamefont{and}
  \bibinfo{author}{\bibfnamefont{J.~S.} \bibnamefont{Read}},
  \bibinfo{journal}{Phys. Rev.} \textbf{\bibinfo{volume}{D81}},
  \bibinfo{pages}{123016} (\bibinfo{year}{2010}).

\bibitem[{\citenamefont{Damour and Nagar}(2009)}]{Damour:2009vw}
\bibinfo{author}{\bibfnamefont{T.}~\bibnamefont{Damour}} \bibnamefont{and}
  \bibinfo{author}{\bibfnamefont{A.}~\bibnamefont{Nagar}},
  \bibinfo{journal}{Phys. Rev.} \textbf{\bibinfo{volume}{D80}},
  \bibinfo{pages}{084035} (\bibinfo{year}{2009}).

\bibitem[{\citenamefont{Postnikov et~al.}(2010)\citenamefont{Postnikov,
  Prakash, and Lattimer}}]{Postnikov:2010yn}
\bibinfo{author}{\bibfnamefont{S.}~\bibnamefont{Postnikov}},
  \bibinfo{author}{\bibfnamefont{M.}~\bibnamefont{Prakash}}, \bibnamefont{and}
  \bibinfo{author}{\bibfnamefont{J.~M.} \bibnamefont{Lattimer}},
  \bibinfo{journal}{Phys. Rev.} \textbf{\bibinfo{volume}{D82}},
  \bibinfo{pages}{024016} (\bibinfo{year}{2010}).

\bibitem[{\citenamefont{Fattoyev et~al.}(2013)\citenamefont{Fattoyev, Carvajal,
  Newton, and Li}}]{Fattoyev:2012uu}
\bibinfo{author}{\bibfnamefont{F.~J.} \bibnamefont{Fattoyev}},
  \bibinfo{author}{\bibfnamefont{J.}~\bibnamefont{Carvajal}},
  \bibinfo{author}{\bibfnamefont{W.~G.} \bibnamefont{Newton}},
  \bibnamefont{and} \bibinfo{author}{\bibfnamefont{B.-A.} \bibnamefont{Li}},
  \bibinfo{journal}{Phys. Rev.} \textbf{\bibinfo{volume}{C87}},
  \bibinfo{pages}{015806} (\bibinfo{year}{2013}).

\bibitem[{\citenamefont{Steiner et~al.}(2015)\citenamefont{Steiner, Gandolfi,
  Fattoyev, and Newton}}]{Steiner:2014pda}
\bibinfo{author}{\bibfnamefont{A.~W.} \bibnamefont{Steiner}},
  \bibinfo{author}{\bibfnamefont{S.}~\bibnamefont{Gandolfi}},
  \bibinfo{author}{\bibfnamefont{F.~J.} \bibnamefont{Fattoyev}},
  \bibnamefont{and} \bibinfo{author}{\bibfnamefont{W.~G.}
  \bibnamefont{Newton}}, \bibinfo{journal}{Phys. Rev.}
  \textbf{\bibinfo{volume}{C91}}, \bibinfo{pages}{015804}
  (\bibinfo{year}{2015}).

\bibitem[{\citenamefont{Damour et~al.}(1992)\citenamefont{Damour, Soffel, and
  Xu}}]{Damour:1991yw}
\bibinfo{author}{\bibfnamefont{T.}~\bibnamefont{Damour}},
  \bibinfo{author}{\bibfnamefont{M.}~\bibnamefont{Soffel}}, \bibnamefont{and}
  \bibinfo{author}{\bibfnamefont{C.}~\bibnamefont{Xu}}, \bibinfo{journal}{Phys.
  Rev.} \textbf{\bibinfo{volume}{D45}}, \bibinfo{pages}{1017}
  (\bibinfo{year}{1992}).

\bibitem[{\citenamefont{Flanagan and Hinderer}(2008)}]{Flanagan:2007ix}
\bibinfo{author}{\bibfnamefont{E.~E.} \bibnamefont{Flanagan}} \bibnamefont{and}
  \bibinfo{author}{\bibfnamefont{T.}~\bibnamefont{Hinderer}},
  \bibinfo{journal}{Phys. Rev.} \textbf{\bibinfo{volume}{D77}},
  \bibinfo{pages}{021502} (\bibinfo{year}{2008}).

\bibitem[{\citenamefont{Binnington and Poisson}(2009)}]{Binnington:2009bb}
\bibinfo{author}{\bibfnamefont{T.}~\bibnamefont{Binnington}} \bibnamefont{and}
  \bibinfo{author}{\bibfnamefont{E.}~\bibnamefont{Poisson}},
  \bibinfo{journal}{Phys. Rev.} \textbf{\bibinfo{volume}{D80}},
  \bibinfo{pages}{084018} (\bibinfo{year}{2009}).

\bibitem[{\citenamefont{Damour et~al.}(2012)\citenamefont{Damour, Nagar, and
  Villain}}]{Damour:2012yf}
\bibinfo{author}{\bibfnamefont{T.}~\bibnamefont{Damour}},
  \bibinfo{author}{\bibfnamefont{A.}~\bibnamefont{Nagar}}, \bibnamefont{and}
  \bibinfo{author}{\bibfnamefont{L.}~\bibnamefont{Villain}},
  \bibinfo{journal}{Phys. Rev.} \textbf{\bibinfo{volume}{D85}},
  \bibinfo{pages}{123007} (\bibinfo{year}{2012}).

\bibitem[{\citenamefont{Abbott et~al.}(2018)\citenamefont{Abbott, Abbott,
  Abbott, Acernese, Ackley, Adams, Adams, Addesso, Adhikari, Adya
  et~al.}}]{LIGO_new}
\bibinfo{author}{\bibfnamefont{B.~P.} \bibnamefont{Abbott}},
  \bibinfo{author}{\bibfnamefont{R.}~\bibnamefont{Abbott}},
  \bibinfo{author}{\bibfnamefont{T.~D.} \bibnamefont{Abbott}},
  \bibinfo{author}{\bibfnamefont{F.}~\bibnamefont{Acernese}},
  \bibinfo{author}{\bibfnamefont{K.}~\bibnamefont{Ackley}},
  \bibinfo{author}{\bibfnamefont{C.}~\bibnamefont{Adams}},
  \bibinfo{author}{\bibfnamefont{T.}~\bibnamefont{Adams}},
  \bibinfo{author}{\bibfnamefont{P.}~\bibnamefont{Addesso}},
  \bibinfo{author}{\bibfnamefont{R.~X.} \bibnamefont{Adhikari}},
  \bibinfo{author}{\bibfnamefont{V.~B.} \bibnamefont{Adya}},
  \bibnamefont{et~al.} (\bibinfo{collaboration}{The LIGO Scientific
  Collaboration and the Virgo Collaboration}), \bibinfo{journal}{Phys. Rev.
  Lett.} \textbf{\bibinfo{volume}{121}}, \bibinfo{pages}{161101}
  (\bibinfo{year}{2018}),
  \urlprefix\url{https://link.aps.org/doi/10.1103/PhysRevLett.121.161101}.

\bibitem[{\citenamefont{Sun et~al.}(2020)\citenamefont{Sun, Wen, and
  Wang}}]{PhysRevD.102.023039}
\bibinfo{author}{\bibfnamefont{W.}~\bibnamefont{Sun}},
  \bibinfo{author}{\bibfnamefont{D.}~\bibnamefont{Wen}}, \bibnamefont{and}
  \bibinfo{author}{\bibfnamefont{J.}~\bibnamefont{Wang}},
  \bibinfo{journal}{Phys. Rev. D} \textbf{\bibinfo{volume}{102}},
  \bibinfo{pages}{023039} (\bibinfo{year}{2020}),
  \urlprefix\url{https://link.aps.org/doi/10.1103/PhysRevD.102.023039}.

\bibitem[{\citenamefont{{Akmal} et~al.}(1998)\citenamefont{{Akmal},
  {Pandharipande}, and {Ravenhall}}}]{Akmal:1998}
\bibinfo{author}{\bibfnamefont{A.}~\bibnamefont{{Akmal}}},
  \bibinfo{author}{\bibfnamefont{V.~R.} \bibnamefont{{Pandharipande}}},
  \bibnamefont{and} \bibinfo{author}{\bibfnamefont{D.~G.}
  \bibnamefont{{Ravenhall}}}, \bibinfo{journal}{\prc}
  \textbf{\bibinfo{volume}{58}}, \bibinfo{pages}{1804} (\bibinfo{year}{1998}),
  \eprint{nucl-th/9804027}.

\bibitem[{\citenamefont{{Potekhin} et~al.}(2013)\citenamefont{{Potekhin},
  {Fantina}, {Chamel}, {Pearson}, and {Goriely}}}]{Potekhin:2013}
\bibinfo{author}{\bibfnamefont{A.~Y.} \bibnamefont{{Potekhin}}},
  \bibinfo{author}{\bibfnamefont{A.~F.} \bibnamefont{{Fantina}}},
  \bibinfo{author}{\bibfnamefont{N.}~\bibnamefont{{Chamel}}},
  \bibinfo{author}{\bibfnamefont{J.~M.} \bibnamefont{{Pearson}}},
  \bibnamefont{and}
  \bibinfo{author}{\bibfnamefont{S.}~\bibnamefont{{Goriely}}},
  \bibinfo{journal}{\aap} \textbf{\bibinfo{volume}{560}}, \bibinfo{eid}{A48}
  (\bibinfo{year}{2013}), \eprint{1310.0049}.

\bibitem[{\citenamefont{{Fattoyev} et~al.}(2020)\citenamefont{{Fattoyev},
  {Horowitz}, {Piekarewicz}, and {Reed}}}]{big-apple:2020}
\bibinfo{author}{\bibfnamefont{F.~J.} \bibnamefont{{Fattoyev}}},
  \bibinfo{author}{\bibfnamefont{C.~J.} \bibnamefont{{Horowitz}}},
  \bibinfo{author}{\bibfnamefont{J.}~\bibnamefont{{Piekarewicz}}},
  \bibnamefont{and} \bibinfo{author}{\bibfnamefont{B.}~\bibnamefont{{Reed}}},
  \bibinfo{journal}{arXiv e-prints} \bibinfo{eid}{arXiv:2007.03799}
  (\bibinfo{year}{2020}), \eprint{2007.03799}.

\bibitem[{\citenamefont{Friedman and Pandharipande}(1981)}]{Friedman:1981qw}
\bibinfo{author}{\bibfnamefont{B.}~\bibnamefont{Friedman}} \bibnamefont{and}
  \bibinfo{author}{\bibfnamefont{V.~R.} \bibnamefont{Pandharipande}},
  \bibinfo{journal}{Nucl. Phys.} \textbf{\bibinfo{volume}{A361}},
  \bibinfo{pages}{502} (\bibinfo{year}{1981}).

\bibitem[{\citenamefont{Chen and Piekarewicz}(2014)}]{Chen:2014sca}
\bibinfo{author}{\bibfnamefont{W.-C.} \bibnamefont{Chen}} \bibnamefont{and}
  \bibinfo{author}{\bibfnamefont{J.}~\bibnamefont{Piekarewicz}},
  \bibinfo{journal}{Phys. Rev.} \textbf{\bibinfo{volume}{C90}},
  \bibinfo{pages}{044305} (\bibinfo{year}{2014}).

\bibitem[{\citenamefont{Tolos et~al.}(2017)\citenamefont{Tolos, Centelles, and
  Ramos}}]{Tolos:2017}
\bibinfo{author}{\bibfnamefont{L.}~\bibnamefont{Tolos}},
  \bibinfo{author}{\bibfnamefont{M.}~\bibnamefont{Centelles}},
  \bibnamefont{and} \bibinfo{author}{\bibfnamefont{A.}~\bibnamefont{Ramos}},
  \bibinfo{journal}{Publications of the Astronomical Society of Australia}
  \textbf{\bibinfo{volume}{34}}, \bibinfo{pages}{e065} (\bibinfo{year}{2017}).

\bibitem[{\citenamefont{Chen and Piekarewicz}(2015)}]{Chen:2014mza}
\bibinfo{author}{\bibfnamefont{W.-C.} \bibnamefont{Chen}} \bibnamefont{and}
  \bibinfo{author}{\bibfnamefont{J.}~\bibnamefont{Piekarewicz}},
  \bibinfo{journal}{Phys. Lett.} \textbf{\bibinfo{volume}{B748}},
  \bibinfo{pages}{284} (\bibinfo{year}{2015}).

\bibitem[{\citenamefont{{Glendenning}}(1985)}]{Glendenning:1985}
\bibinfo{author}{\bibfnamefont{N.~K.} \bibnamefont{{Glendenning}}},
  \bibinfo{journal}{\apj} \textbf{\bibinfo{volume}{293}}, \bibinfo{pages}{470}
  (\bibinfo{year}{1985}).

\bibitem[{\citenamefont{Fattoyev et~al.}(2010)\citenamefont{Fattoyev, Horowitz,
  Piekarewicz, and Shen}}]{Fattoyev:2010mx}
\bibinfo{author}{\bibfnamefont{F.~J.} \bibnamefont{Fattoyev}},
  \bibinfo{author}{\bibfnamefont{C.~J.} \bibnamefont{Horowitz}},
  \bibinfo{author}{\bibfnamefont{J.}~\bibnamefont{Piekarewicz}},
  \bibnamefont{and} \bibinfo{author}{\bibfnamefont{G.}~\bibnamefont{Shen}},
  \bibinfo{journal}{Phys. Rev.} \textbf{\bibinfo{volume}{C82}},
  \bibinfo{pages}{055803} (\bibinfo{year}{2010}).

\bibitem[{\citenamefont{{Nandi} et~al.}(2019)\citenamefont{{Nandi}, {Char}, and
  {Pal}}}]{Nandi:2019}
\bibinfo{author}{\bibfnamefont{R.}~\bibnamefont{{Nandi}}},
  \bibinfo{author}{\bibfnamefont{P.}~\bibnamefont{{Char}}}, \bibnamefont{and}
  \bibinfo{author}{\bibfnamefont{S.}~\bibnamefont{{Pal}}},
  \bibinfo{journal}{\prc} \textbf{\bibinfo{volume}{99}}, \bibinfo{eid}{052802}
  (\bibinfo{year}{2019}), \eprint{1809.07108}.

\bibitem[{\citenamefont{{M{\"u}ther} et~al.}(1987)\citenamefont{{M{\"u}ther},
  {Prakash}, and {Ainsworth}}}]{Muther:1987}
\bibinfo{author}{\bibfnamefont{H.}~\bibnamefont{{M{\"u}ther}}},
  \bibinfo{author}{\bibfnamefont{M.}~\bibnamefont{{Prakash}}},
  \bibnamefont{and} \bibinfo{author}{\bibfnamefont{T.~L.}
  \bibnamefont{{Ainsworth}}}, \bibinfo{journal}{Physics Letters B}
  \textbf{\bibinfo{volume}{199}}, \bibinfo{pages}{469} (\bibinfo{year}{1987}).

\bibitem[{\citenamefont{Lalazissis et~al.}(1997)\citenamefont{Lalazissis,
  Konig, and Ring}}]{Lalazissis:1996rd}
\bibinfo{author}{\bibfnamefont{G.~A.} \bibnamefont{Lalazissis}},
  \bibinfo{author}{\bibfnamefont{J.}~\bibnamefont{Konig}}, \bibnamefont{and}
  \bibinfo{author}{\bibfnamefont{P.}~\bibnamefont{Ring}},
  \bibinfo{journal}{Phys. Rev.} \textbf{\bibinfo{volume}{C55}},
  \bibinfo{pages}{540} (\bibinfo{year}{1997}).

\bibitem[{\citenamefont{Carriere et~al.}(2003)\citenamefont{Carriere, Horowitz,
  and Piekarewicz}}]{Carriere:2002bx}
\bibinfo{author}{\bibfnamefont{J.}~\bibnamefont{Carriere}},
  \bibinfo{author}{\bibfnamefont{C.~J.} \bibnamefont{Horowitz}},
  \bibnamefont{and}
  \bibinfo{author}{\bibfnamefont{J.}~\bibnamefont{Piekarewicz}},
  \bibinfo{journal}{Astrophys. J.} \textbf{\bibinfo{volume}{593}},
  \bibinfo{pages}{463} (\bibinfo{year}{2003}).

\bibitem[{\citenamefont{Douchin and Haensel}(2000)}]{Douchin:2000kx}
\bibinfo{author}{\bibfnamefont{F.}~\bibnamefont{Douchin}} \bibnamefont{and}
  \bibinfo{author}{\bibfnamefont{P.}~\bibnamefont{Haensel}},
  \bibinfo{journal}{Phys. Lett.} \textbf{\bibinfo{volume}{B485}},
  \bibinfo{pages}{107} (\bibinfo{year}{2000}), \eprint{astro-ph/0006135}.

\bibitem[{\citenamefont{Fattoyev and Piekarewicz}(2013)}]{Fattoyev:2013yaa}
\bibinfo{author}{\bibfnamefont{F.~J.} \bibnamefont{Fattoyev}} \bibnamefont{and}
  \bibinfo{author}{\bibfnamefont{J.}~\bibnamefont{Piekarewicz}},
  \bibinfo{journal}{Phys. Rev. Lett.} \textbf{\bibinfo{volume}{111}},
  \bibinfo{pages}{162501} (\bibinfo{year}{2013}).

\bibitem[{\citenamefont{Wiringa et~al.}(1988)\citenamefont{Wiringa, Fiks, and
  Fabrocini}}]{Wiringa:1988tp}
\bibinfo{author}{\bibfnamefont{R.~B.} \bibnamefont{Wiringa}},
  \bibinfo{author}{\bibfnamefont{V.}~\bibnamefont{Fiks}}, \bibnamefont{and}
  \bibinfo{author}{\bibfnamefont{A.}~\bibnamefont{Fabrocini}},
  \bibinfo{journal}{Phys. Rev.} \textbf{\bibinfo{volume}{C38}},
  \bibinfo{pages}{1010} (\bibinfo{year}{1988}).

\bibitem[{\citenamefont{Duflo and Zuker}(1995)}]{Duflo:1995}
\bibinfo{author}{\bibfnamefont{J.}~\bibnamefont{Duflo}} \bibnamefont{and}
  \bibinfo{author}{\bibfnamefont{A.}~\bibnamefont{Zuker}},
  \bibinfo{journal}{Phys. Rev.} \textbf{\bibinfo{volume}{C52}},
  \bibinfo{pages}{R23} (\bibinfo{year}{1995}).

\bibitem[{\citenamefont{{Piekarewicz} and {Fattoyev}}(2019)}]{piekarewicz:2019}
\bibinfo{author}{\bibfnamefont{J.}~\bibnamefont{{Piekarewicz}}}
  \bibnamefont{and} \bibinfo{author}{\bibfnamefont{F.~J.}
  \bibnamefont{{Fattoyev}}}, \bibinfo{journal}{\prc}
  \textbf{\bibinfo{volume}{99}}, \bibinfo{eid}{045802} (\bibinfo{year}{2019}),
  \eprint{1812.09974}.

\bibitem[{Oze()}]{Ozel}
\urlprefix\url{http://xtreme.as.arizona.edu/neutronstars/}.

\bibitem[{\citenamefont{Oppenheimer and Volkoff}(1939)}]{Opp39_PR55}
\bibinfo{author}{\bibfnamefont{J.~R.} \bibnamefont{Oppenheimer}}
  \bibnamefont{and} \bibinfo{author}{\bibfnamefont{G.~M.}
  \bibnamefont{Volkoff}}, \bibinfo{journal}{Phys. Rev.}
  \textbf{\bibinfo{volume}{55}}, \bibinfo{pages}{374} (\bibinfo{year}{1939}).

\bibitem[{\citenamefont{Glendenning}(2000)}]{Glendenning:2000}
\bibinfo{author}{\bibfnamefont{N.~K.} \bibnamefont{Glendenning}},
  \emph{\bibinfo{title}{Compact Stars}} (\bibinfo{publisher}{Springer-Verlag
  New York}, \bibinfo{year}{2000}).

\bibitem[{\citenamefont{Lattimer and Prakash}(2007)}]{Lattimer:2006xb}
\bibinfo{author}{\bibfnamefont{J.~M.} \bibnamefont{Lattimer}} \bibnamefont{and}
  \bibinfo{author}{\bibfnamefont{M.}~\bibnamefont{Prakash}},
  \bibinfo{journal}{Phys. Rept.} \textbf{\bibinfo{volume}{442}},
  \bibinfo{pages}{109} (\bibinfo{year}{2007}).

\bibitem[{\citenamefont{Lattimer and Prakash}(2001)}]{Lattimer:2000nx}
\bibinfo{author}{\bibfnamefont{J.~M.} \bibnamefont{Lattimer}} \bibnamefont{and}
  \bibinfo{author}{\bibfnamefont{M.}~\bibnamefont{Prakash}},
  \bibinfo{journal}{Astrophys. J.} \textbf{\bibinfo{volume}{550}},
  \bibinfo{pages}{426} (\bibinfo{year}{2001}), \eprint{astro-ph/0002232}.

\bibitem[{\citenamefont{Weinberg}(1972)}]{Weinberg:1972}
\bibinfo{author}{\bibfnamefont{S.}~\bibnamefont{Weinberg}},
  \emph{\bibinfo{title}{Gravitation and Cosmology}} (\bibinfo{publisher}{John
  Wiley \& Sons, New York}, \bibinfo{year}{1972}).

\bibitem[{\citenamefont{{Prakash} et~al.}(1997)\citenamefont{{Prakash},
  {Bombaci}, {Prakash}, {Ellis}, {Lattimer}, and {Knorren}}}]{Prakash:1997}
\bibinfo{author}{\bibfnamefont{M.}~\bibnamefont{{Prakash}}},
  \bibinfo{author}{\bibfnamefont{I.}~\bibnamefont{{Bombaci}}},
  \bibinfo{author}{\bibfnamefont{M.}~\bibnamefont{{Prakash}}},
  \bibinfo{author}{\bibfnamefont{P.~J.} \bibnamefont{{Ellis}}},
  \bibinfo{author}{\bibfnamefont{J.~M.} \bibnamefont{{Lattimer}}},
  \bibnamefont{and}
  \bibinfo{author}{\bibfnamefont{R.}~\bibnamefont{{Knorren}}},
  \bibinfo{journal}{\physrep} \textbf{\bibinfo{volume}{280}},
  \bibinfo{pages}{1} (\bibinfo{year}{1997}), \eprint{nucl-th/9603042}.

\bibitem[{\citenamefont{{Virtanen} et~al.}(2020)\citenamefont{{Virtanen},
  {Gommers}, {Oliphant}, {Haberland}, {Reddy}, {Cournapeau}, {Burovski},
  {Peterson}, {Weckesser}, {Bright} et~al.}}]{2020SciPy-NMeth}
\bibinfo{author}{\bibfnamefont{P.}~\bibnamefont{{Virtanen}}},
  \bibinfo{author}{\bibfnamefont{R.}~\bibnamefont{{Gommers}}},
  \bibinfo{author}{\bibfnamefont{T.~E.} \bibnamefont{{Oliphant}}},
  \bibinfo{author}{\bibfnamefont{M.}~\bibnamefont{{Haberland}}},
  \bibinfo{author}{\bibfnamefont{T.}~\bibnamefont{{Reddy}}},
  \bibinfo{author}{\bibfnamefont{D.}~\bibnamefont{{Cournapeau}}},
  \bibinfo{author}{\bibfnamefont{E.}~\bibnamefont{{Burovski}}},
  \bibinfo{author}{\bibfnamefont{P.}~\bibnamefont{{Peterson}}},
  \bibinfo{author}{\bibfnamefont{W.}~\bibnamefont{{Weckesser}}},
  \bibinfo{author}{\bibfnamefont{J.}~\bibnamefont{{Bright}}},
  \bibnamefont{et~al.}, \bibinfo{journal}{Nature Methods}
  \textbf{\bibinfo{volume}{17}}, \bibinfo{pages}{261} (\bibinfo{year}{2020}).

\bibitem[{cur()}]{curve_fit}
\urlprefix\url{https://docs.scipy.org/doc/scipy/reference/generated/scipy.optimize.curve_fit.html}.

\bibitem[{\citenamefont{Martinez et~al.}(2015)\citenamefont{Martinez, Stovall,
  Freire, Deneva, Jenet, McLaughlin, Bagchi, Bates, and
  Ridolfi}}]{Martinez_2015}
\bibinfo{author}{\bibfnamefont{J.~G.} \bibnamefont{Martinez}},
  \bibinfo{author}{\bibfnamefont{K.}~\bibnamefont{Stovall}},
  \bibinfo{author}{\bibfnamefont{P.~C.~C.} \bibnamefont{Freire}},
  \bibinfo{author}{\bibfnamefont{J.~S.} \bibnamefont{Deneva}},
  \bibinfo{author}{\bibfnamefont{F.~A.} \bibnamefont{Jenet}},
  \bibinfo{author}{\bibfnamefont{M.~A.} \bibnamefont{McLaughlin}},
  \bibinfo{author}{\bibfnamefont{M.}~\bibnamefont{Bagchi}},
  \bibinfo{author}{\bibfnamefont{S.~D.} \bibnamefont{Bates}}, \bibnamefont{and}
  \bibinfo{author}{\bibfnamefont{A.}~\bibnamefont{Ridolfi}},
  \bibinfo{journal}{The Astrophysical Journal} \textbf{\bibinfo{volume}{812}},
  \bibinfo{pages}{143} (\bibinfo{year}{2015}), ISSN \bibinfo{issn}{1538-4357},
  \urlprefix\url{http://dx.doi.org/10.1088/0004-637x/812/2/143}.

\bibitem[{\citenamefont{{Suwa} et~al.}(2018)\citenamefont{{Suwa}, {Yoshida},
  {Shibata}, {Umeda}, and {Takahashi}}}]{suwa:2018}
\bibinfo{author}{\bibfnamefont{Y.}~\bibnamefont{{Suwa}}},
  \bibinfo{author}{\bibfnamefont{T.}~\bibnamefont{{Yoshida}}},
  \bibinfo{author}{\bibfnamefont{M.}~\bibnamefont{{Shibata}}},
  \bibinfo{author}{\bibfnamefont{H.}~\bibnamefont{{Umeda}}}, \bibnamefont{and}
  \bibinfo{author}{\bibfnamefont{K.}~\bibnamefont{{Takahashi}}},
  \bibinfo{journal}{\mnras} \textbf{\bibinfo{volume}{481}},
  \bibinfo{pages}{3305} (\bibinfo{year}{2018}), \eprint{1808.02328}.

\bibitem[{\citenamefont{Abbott et~al.}(2020)\citenamefont{Abbott, Abbott,
  Abraham, Acernese, Ackley, Adams, Adhikari, Adya, Affeldt, Agathos
  et~al.}}]{Abbott_2020}
\bibinfo{author}{\bibfnamefont{R.}~\bibnamefont{Abbott}},
  \bibinfo{author}{\bibfnamefont{T.~D.} \bibnamefont{Abbott}},
  \bibinfo{author}{\bibfnamefont{S.}~\bibnamefont{Abraham}},
  \bibinfo{author}{\bibfnamefont{F.}~\bibnamefont{Acernese}},
  \bibinfo{author}{\bibfnamefont{K.}~\bibnamefont{Ackley}},
  \bibinfo{author}{\bibfnamefont{C.}~\bibnamefont{Adams}},
  \bibinfo{author}{\bibfnamefont{R.~X.} \bibnamefont{Adhikari}},
  \bibinfo{author}{\bibfnamefont{V.~B.} \bibnamefont{Adya}},
  \bibinfo{author}{\bibfnamefont{C.}~\bibnamefont{Affeldt}},
  \bibinfo{author}{\bibfnamefont{M.}~\bibnamefont{Agathos}},
  \bibnamefont{et~al.}, \bibinfo{journal}{The Astrophysical Journal}
  \textbf{\bibinfo{volume}{896}}, \bibinfo{pages}{L44} (\bibinfo{year}{2020}),
  ISSN \bibinfo{issn}{2041-8213},
  \urlprefix\url{http://dx.doi.org/10.3847/2041-8213/ab960f}.

\bibitem[{\citenamefont{{Tews} et~al.}(2020)\citenamefont{{Tews}, {Pang},
  {Dietrich}, {Coughlin}, {Antier}, {Bulla}, {Heinzel}, and
  {Issa}}}]{Tews:2020}
\bibinfo{author}{\bibfnamefont{I.}~\bibnamefont{{Tews}}},
  \bibinfo{author}{\bibfnamefont{P.~T.~H.} \bibnamefont{{Pang}}},
  \bibinfo{author}{\bibfnamefont{T.}~\bibnamefont{{Dietrich}}},
  \bibinfo{author}{\bibfnamefont{M.~W.} \bibnamefont{{Coughlin}}},
  \bibinfo{author}{\bibfnamefont{S.}~\bibnamefont{{Antier}}},
  \bibinfo{author}{\bibfnamefont{M.}~\bibnamefont{{Bulla}}},
  \bibinfo{author}{\bibfnamefont{J.}~\bibnamefont{{Heinzel}}},
  \bibnamefont{and} \bibinfo{author}{\bibfnamefont{L.}~\bibnamefont{{Issa}}},
  \bibinfo{journal}{arXiv e-prints} \bibinfo{eid}{arXiv:2007.06057}
  (\bibinfo{year}{2020}), \eprint{2007.06057}.

\bibitem[{\citenamefont{{Lim} et~al.}(2020)\citenamefont{{Lim}, {Bhattacharya},
  {Holt}, and {Pati}}}]{Lim:2020}
\bibinfo{author}{\bibfnamefont{Y.}~\bibnamefont{{Lim}}},
  \bibinfo{author}{\bibfnamefont{A.}~\bibnamefont{{Bhattacharya}}},
  \bibinfo{author}{\bibfnamefont{J.~W.} \bibnamefont{{Holt}}},
  \bibnamefont{and} \bibinfo{author}{\bibfnamefont{D.}~\bibnamefont{{Pati}}},
  \bibinfo{journal}{arXiv e-prints} \bibinfo{eid}{arXiv:2007.06526}
  (\bibinfo{year}{2020}), \eprint{2007.06526}.

\bibitem[{\citenamefont{{Dexheimer} et~al.}(2020)\citenamefont{{Dexheimer},
  {Gomes}, {Kl{\"a}hn}, {Han}, and {Salinas}}}]{Dexheimer:2020}
\bibinfo{author}{\bibfnamefont{V.}~\bibnamefont{{Dexheimer}}},
  \bibinfo{author}{\bibfnamefont{R.~O.} \bibnamefont{{Gomes}}},
  \bibinfo{author}{\bibfnamefont{T.}~\bibnamefont{{Kl{\"a}hn}}},
  \bibinfo{author}{\bibfnamefont{S.}~\bibnamefont{{Han}}}, \bibnamefont{and}
  \bibinfo{author}{\bibfnamefont{M.}~\bibnamefont{{Salinas}}},
  \bibinfo{journal}{arXiv e-prints} \bibinfo{eid}{arXiv:2007.08493}
  (\bibinfo{year}{2020}), \eprint{2007.08493}.

\bibitem[{\citenamefont{{Roupas}}(2020)}]{Roupas:2020}
\bibinfo{author}{\bibfnamefont{Z.}~\bibnamefont{{Roupas}}},
  \bibinfo{journal}{arXiv e-prints} \bibinfo{eid}{arXiv:2007.10679}
  (\bibinfo{year}{2020}), \eprint{2007.10679}.

\end{thebibliography}


\nocite{*}
\end{document}